\documentstyle[amssymb,aps,prl,graphicx,preprint]{revtex}

\begin{document}

\title{Stationary and Oscillatory Spatial Patterns Induced by
Global Periodic Switching}
\author{J. Buceta$^{1}$, Katja Lindenberg$^{1}$, and J.M.R. Parrondo$^{2}$}
\address{$^{1}$Department of Chemistry and Biochemistry and Institute for Nonlinear
Science,\\ University of California San Diego, 9500 Gilman Drive, La Jolla,
CA 92093-0340, USA\\
$^{2}$Departamento de F\'{\i}sica At\'{o}mica, Molecular y Nuclear,\\
Universidad
Complutense de Madrid, Ciudad Universitaria s/n, 28040 Madrid, Spain}
\date{\today}
\maketitle

\begin{abstract}
We propose a new mechanism for pattern formation based on the global
alternation of two dynamics neither of which exhibits patterns.
When driven by either one of the separate dynamics, the system goes
to a spatially homogeneous state associated with that dynamics. 
However, when the two dynamics are globally alternated sufficiently
rapidly, the system exhibits stationary spatial patterns.  Somewhat
slower switching leads to oscillatory patterns.  We support
our findings by numerical simulations and discuss the results in terms
of the symmetries of the system and the ratio of two relevant characteristic
times, the switching period and the relaxation time to a homogeneous
state in each separate dynamics.
\end{abstract}

Patterns in non-equilibrium systems arise in a variety of ways. 
For example, so-called dissipative structures arise from perturbations
of homogeneous systems if the input of energy is properly balanced by
dissipation~\cite{general}  Perhaps the best-known dissipative structure
is the Ben\'{a}rd instability, formed when a layer of liquid is heated from
below. At a given temperature, heat conduction starts to occur
predominantly through convection, and regularly spaced, hexagonal
convection cells are formed in the layer of liquid. This structure is
present only as long as there is a supply of heat and disappears
when this ceases.  Many other examples involve chemical
oscillations in dissipative open systems~\cite{general}.  Patterns may also
form by the temporal modulation of a parameter in systems
that undergo Hopf bifurcations.  This modulation may stabilitze
standing waves or wave patterns that are otherwise unstable for any
constant value of the parameter in this regime~\cite{SHforced1,sagues}. 
A different
sort of phenomenon, that of noise-induced pattern formation, may occur
when an external spatially and temporally uncorrelated noise affects a
system parameter.  The parameter fluctuations may induce spatial
organization and pattern formation in a system where only a
homogeneous state is stable when the parameter is
constant~\cite{chris,juanma}.

In this Letter we identify and illustrate a new mechanism for
pattern formation induced by a global periodic alternation between
two dynamics, each of which by itself leads to a (different)
spatially homogeneous state.  This mechanism is inspired by a number
of examples where simple periodic switching between
two dynamics, each of which produces ``uninteresting" or ``disordered"
or even ``undesirable" states, leads to an ``interesting" or ``ordered" or
``desirable" outcome.  One example is the flashing Brownian ratchet, a
rectifier of thermal fluctuations that can induce directed motion of
Brownian particles merely by turning the ratchet potential on and off
periodically~\cite{adjari}.  Another (in turn inspired by
the ratchet example) is that
of so-called paradoxical games, where the alternation of two losing games
may lead to a winning game~\cite{abbot}.  Our proposed
mechanism applies these ideas to spatially extended systems.

We focus on a class of model systems based on the
Swift-Hohenberg (SH) equation~\cite{swifthohenberg}: 
\begin{equation}
\stackrel{\cdot }{\varphi }\left( {\bf r},t\right) =-V^{^{\prime }}\left(
\varphi \left( {\bf r},t\right) \right) +{\cal L}\varphi \left( {\bf r}
,t\right) +\xi \left( {\bf r},t\right)\text{,}  \label{sh0}
\end{equation}
where the operator ${\cal L}=-\left( 1+\nabla ^{2}\right) ^{2}$ is a spatial
coupling,
and the white zero-centered Gaussian noise $\xi \left( {\bf r},t\right)$
with correlation 
$\left\langle \xi \left( {\bf r},t\right) \langle \xi \left( {{\bf r}
^{\prime }},t^{\prime }\right) \right\rangle =\sigma ^{2}\delta (t-t^{\prime
})\delta ({\bf r}-{\bf r}^{\prime })$ accounts for thermal
or other fluctuations.
The dynamics of the system is determined by the local potential
$V(\varphi )$, which in the standard SH model is an even quartic
function of $\varphi$.
The SH  model leads to pattern formation
(e.g. to the appearance of Rayleigh-Ben\'{a}rd convective rolls) 
when the local potential has two stable equilibrium
points.  However, if $V(\varphi)$ is {\em monostable}, no spatial
structures appear in
this system and the steady state is spatially homogeneous.  The stability
boundary is identified by determining a uniform solution of the (noiseless) 
evolution equation (\ref{sh0}) and linearizing about this solution.  The
stationarity condition
$V^\prime (\varphi) +\varphi=0$
(the term $\varphi$ arises from the $1$ in the coupling operator) has the
solution $\varphi\equiv \tilde{\varphi}$. Setting $\varphi=\tilde{\varphi}
+\Delta\varphi$ gives for the Fourier transform (denoted by a hat)
of the linearized equation 
\begin{equation}
\Delta \dot{\widehat{\varphi}}({\mathbf k},t)=-\left[ V^{\prime \prime}
(\tilde{\varphi})+(1-\left| \mathbf{k}\right| ^{2})^{2}\right] \Delta
\widehat{\varphi}(\mathbf{k},t), 
\label{linear3}
\end{equation}
and if $V^{\prime \prime }(\tilde{\varphi})<0$ a morphological instability
first sets in for the modes with $|{\bf k}|=1$.
Note that in the usual SH model the even local potential leads to
$\tilde{\varphi}=0$, and that in general
the patterns that emerge depend on the
symmetry of the potential.  Thus, symmetry of the evolution equation
under the $\varphi \longleftrightarrow -\varphi$ inversion leads to roll-shaped
patterns, while absence of this symmetry leads to 
hexagonally-arranged spot-like structures~\cite{cross-hohenberg1}.

Consider now a {\em global} periodic switching of period $T$ between
two local
potentials, $V_{1}$ and $V_{2}$.  Thus, {\em every}
point is subject to the same local potential at any given time.
The system is governed by the equation
\begin{equation}
\stackrel{\cdot }{\varphi }\left( {\bf r},t\right) =-V_{+}^{^{\prime
}}\left( \varphi \left( {\bf r},t\right) \right) -\mu \left( t\right)
V_{-}^{^{\prime }}\left( \varphi \left( {\bf r},t\right) \right) +{\cal L}
\varphi \left( {\bf r},t\right) +\xi \left( {\bf r},t\right)\text{,}
\label{4}
\end{equation}
\noindent
where $V_{\pm }\left( \varphi \right) \equiv
\left( V_{1}\left( \varphi \right)
\pm V_{2}\left( \varphi \right) \right) /2$ and
$\mu \left(t\right)$
takes on the value $1$ if $t~{\rm mod}~T<T/2$ and $-1$ if $t~{\rm 
mod}~T>T/2$.
The resulting equation is a SH
model with dichotomous periodic external forcing. To the best
of our knowledge, this
forcing differs from those previously considered in the
literature~\cite{SHforced1}.   

In order to investigate whether the switching mechanism can lead to pattern
formation, we focus on local potentials $V_1(\varphi)$
and $V_2(\varphi)$ that satisfy the stability conditions,
\begin{equation}
V_i^\prime (\tilde{\varphi}_i) +\tilde{\varphi}_i=0  \ \ \mbox{and} \ \
V_i^{\prime\prime}(\tilde{\varphi}_i) > 0,
\label{cond}
\end{equation}
(monostable potentials)
whereas their average 
$V_+(\varphi)$ does not.
Note that
the condition that $V_+(\varphi)$ {\em not} satisfy (\ref{cond}) implies
that both local potentials can not be quadratic. As a consequence, for
the proposed pattern formation mechanism non-linearity is a necessary
component.

When each potential acts
separately, the system tends to a homogeneous state.
The expectation of pattern formation when the potentials are periodically
alternated arises 
as follows.  If the switching period $T$ is large, every point in the
system has time to equilibrate to the local potential before
the local potentials switch.  Regardless of the initial distribution,
the entire system is expected to oscillate with period $T$
between the homogeneous states corresponding to each potential.
On the other hand, if $T$ is sufficiently
small (see below), $\mu(t)$ can be adiabatically eliminated and 
replaced by its average value, which is zero. The
system is then driven by the potential $V_+(\varphi)$, for which there is at
least one state $\tilde{\varphi}$ with
\begin{equation}
V_+^\prime (\tilde{\varphi}) +\tilde{\varphi}=0  \ \ \mbox{and} \ \
V_+^{\prime\prime}(\tilde{\varphi}) < 0,
\label{condinst}
\end{equation}
and thus patterns appropriate to this local potential are expected to
occur.  Away from these extremes, the behavior depends on the
switching rate.  More specifically,
the crossover time
$t_r$ between slow and fast switching is  
the smaller of $t_{1\to 2}$ and $t_{2\to 1}$,
where $t_{i\to j}$ is the relaxation time, {\em under the action of}
$V_j$, of the homogeneous state associated with $V_i$.
We can estimate $t_{i\to j}$ by
focusing only on the ${\bf k}=0$ mode and assuming that,
when the potential switches from $V_i$ to $V_j$, the mode amplitude
behaves as a Brownian particle initially equilibrated in the
effective local potential
$\widetilde{V}_{i}\left( \varphi \right) = V_{i}\left( \varphi \right)
+\frac{\varphi ^{2}}{ 2}$.  When the local potential is switched,
this point, which up to that moment was stable,
becomes unstable. The relaxation
time to the new homogeneous state associated with $V_j$ is
the time that it takes the Brownian particle to roll down the potential
hill to the new equilibrium point~\cite{malakhov}:
\begin{equation}
t_{i\rightarrow j}=\frac{2}{\sigma ^{2}}
\int_{\tilde{\varphi}_{i}}^{\tilde{\varphi}_{j}}
{\rm d}y
\exp\left(\frac{2}{\sigma ^{2}}\widetilde{V}_{j}
(y)\right) 
\int_{\tilde{\varphi}_{i}}^{y} {\rm d}z\exp\left(-\frac{2}
{\sigma ^{2}}\widetilde{V}_{j}( z)\right). 
\label{tij}
\end{equation}
The behavior of the system (\ref{4}) can be characterized by the ratio 
of the time $T/2$ that the system spends in each dynamics to the
crossover time $t_r$, $r\equiv T/2t_r$.  

\begin{figure}
\begin{center}
\includegraphics[width=6cm]{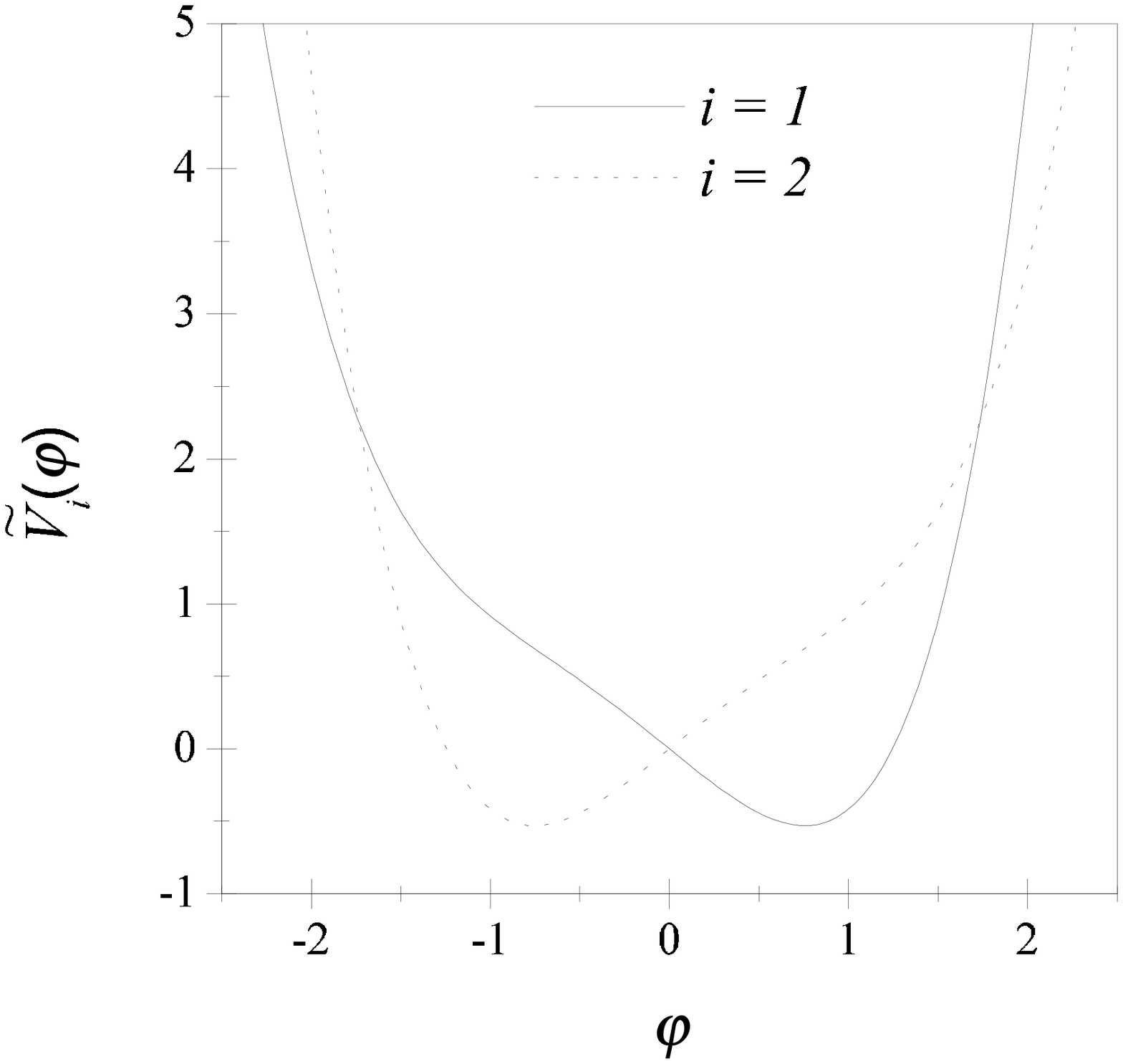}
\end{center}
\caption{Effective local potentials $\widetilde{V}_1(\varphi)$ (solid
curve) and $\widetilde{V}_2(\varphi)$ (dotted curve) with 
$A_{1}=A_{2}=1$.  The mirror symmetry is broken if $A_1\neq A_2$.
}
\label{figpotentials}
\end{figure}

Let us focus on a particular choice of potentials satisfying the
conditions (\ref{cond}) and (\ref{condinst}): 
\begin{equation}
V_{1,2}\left( \varphi \right) =A_{1,2}\left( \frac{\varphi ^{4}}{4}\pm \frac{
\varphi ^{3}}{3}-\frac{\varphi ^{2}}{2}\mp \varphi \right) \text{,}
\label{potentials}
\end{equation}
where $A_{1,2}$ are positive constants.  The
corresponding effective local potentials with $A_1=A_2=1$
are shown in Fig.~\ref{figpotentials}.  The average potential is 
\begin{equation}
V_{+}\left( \varphi \right) =a_{+}\frac{\varphi ^{4}}{4}+a_{-}\frac{\varphi
^{3}}{3}-a_{+}\frac{\varphi ^{2}}{2}-a_{-}\varphi  \label{vmas}
\end{equation}
where $a_{\pm}=\left( A_{1}\pm A_{2}\right) /2$.
If $r\gg 1$ the entire system
alternates between the homogeneous states $\widetilde{\varphi }_{1,2}$.
When $r\lesssim 1$, where we
expect pattern formation with outcomes
dependent on the specific value of $r$ and on the equality or
inequality of the parameters $A_{1,2}$.
We support our reasoning with $2d$ numerical simulations of a discretized
form of Eq.~(\ref{4}) with these potentials, periodic boundary
conditions, and parameters $L_{x}=L_{y}=64$,
$\sigma =10^{-2}$ (the fluctuations must be sufficiently small not to
swamp the potential barrier in $V_+$). Since the most unstable
mode is $\left| {\bf k}^{\ast
}\right| \simeq 1$ \cite{juanma}, the typical wavelength of any pattern is
expected to be $\lambda =2\pi /\left| {\bf k}^{\ast }\right|
\simeq 2\pi $ and the aspect ratio $L/\lambda \sim 10$.
In our simulations we either take the initial field to be random according
to a Gaussian distribution (in which case the additive fluctuations can
actually be omitted entirely), or we can take an arbitrary initial
condition (e.g. all points equilibrated with $V_1$), in which case the
fluctuations will distribute the field in any case.  In some cases we
choose an initial configuration that facilitates arrival at a particular
final state simply to avoid a very long simulation time.  At the initial
time every local potential is set to, say, $V_1$, at time $T/2$ all the
potentials are switched to $V_2$, and so on.  

\begin{figure}
\begin{center}
\includegraphics[width=4.8cm]{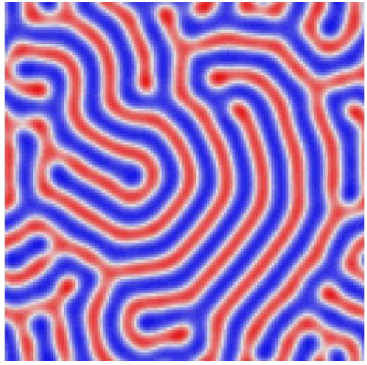}
\end{center}
\caption{Density plot of the field for the case $A_{1}=A_{2}=1$ and $r=0.25$.
The roll-shaped pattern is essentially (see text) stationary.} 
\label{figrolls}
\end{figure}

If $A_{1}=A_{2}$, $V_{+}\left( \varphi \right)$ is an even
function and, in the limit $r\rightarrow 0$, Eq. (\ref{4}) satisfies
inversion symmetry under the transformation
$\varphi \longleftrightarrow -\varphi $. In this
case we expect the appearance of {\em stationary rolls} \cite{cross-hohenberg1}.  This is clearly
seen in Fig.~\ref{figrolls}.
Note that the widths of the rolls are consistent with the aspect ratio given
earlier.  On an
extremely long time scale, as in the SH model, the rolls line
up in a more ordered fashion. 

\begin{figure}
\begin{center}
\includegraphics[width=4.8cm]{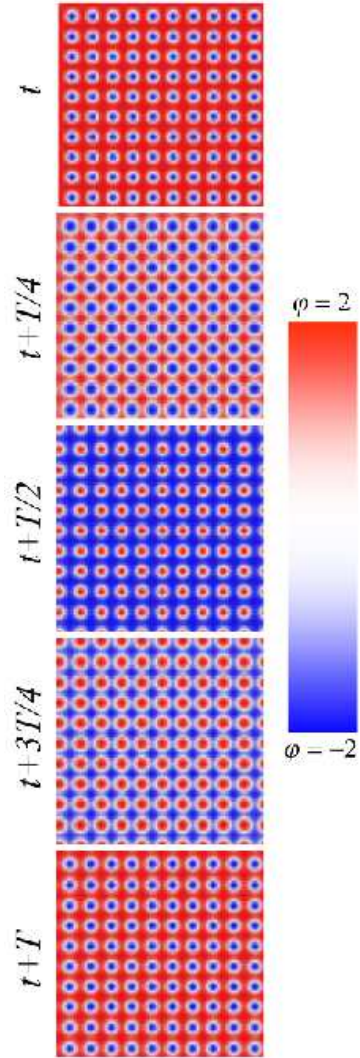}
\end{center}
\caption{Snapshots of the field during a full period of the dichotomous
forcing ($A_{1}=A_{2}=1$ and $r=1.15$). Note the oscillations of
the field between the
square lattice and its glide-transformed one.}
\label{figsquares}
\end{figure}

As $r$ increases toward the ``resonance" condition
$r\thickapprox 1$, the contribution of
$V_{-}^{^{\prime }}\left( \varphi \right)$ can no longer be neglected.
Hence Eq.~(\ref {4}) will lack the symmetry
$\varphi \longleftrightarrow -\varphi $, and an 
{\em oscillatory} spot-like pattern is expected \cite{cross-hohenberg1}. Furthermore,
with A$_{1}=A_{2}$, Eq.~(\ref{4}) is invariant under the combined
transformation $\left\{ \varphi \longleftrightarrow -\varphi ,\mu
\longleftrightarrow -\mu \right\}$, which requires a {\em square} spatial
arrangement of the oscillatory pattern \cite{cross-hohenberg1}.  A realization is shown in
Fig.~\ref{figsquares}, for which 
$t_{r} = t_{1\rightarrow 2}=t_{2\rightarrow 1}\thickapprox 2$.
The resonant period of the forcing is $T\thickapprox 4$. 
We show color encoded snapshots of the
field, where the oscillatory square patterns are
clearly visible. Note that the field oscillates between a square lattice and
its glide-transformed one.   In other words, the spot centers do not move or
oscillate; it is the surrounding background that oscillates. 
We are able to reproduce this behavior analytically through a
decomposition in a small number of modes~\cite{next}.  
The size of the pattern units is again consistent with the aspect
ratio given earlier.
The oscillatory pattern and
the lattice arrangement of the spots resemble the so-called
{\em oscillons} found in vibrating granular media and clay \cite{swinney}.

\begin{figure}
\begin{center}
\includegraphics[width=4.8cm]{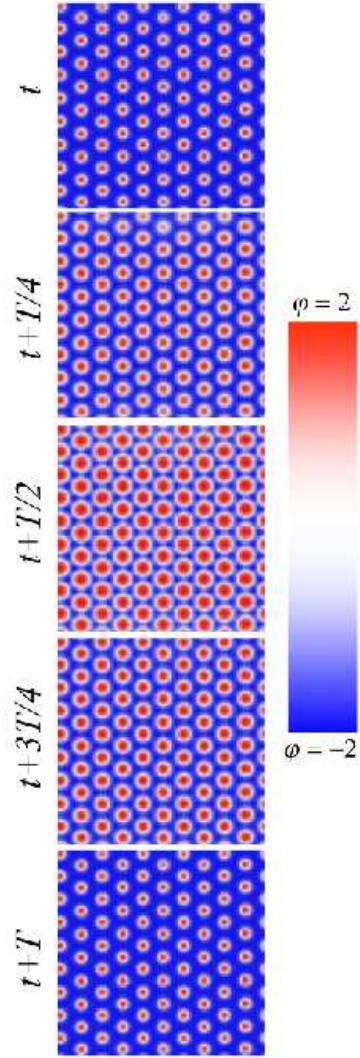}
\end{center}
\caption{Snapshots of the field for the case $A_{1}=1, A_{2}=2$
and $r=0.94$ during a full period of the forcing
function. The localized excitations are arranged in a
hexagonal lattice.}
\label{fighexagons}
\end{figure}

The behavior is in some ways simpler when $A_{1}\neq A_{2}$.
The inversion symmetry $\varphi \longleftrightarrow -\varphi $ is
no longer satisfied.  Therefore, both stationary ($r\ll 1$) and
dynamic ($r\approx 1$) hexagonal spot patterns are expected \cite{cross-hohenberg1}.
We have simulated the case $A_{1}=1$, $A_{2}=2$, for which the 
relaxation time is calculated to be $t_{r}\thickapprox 1.6$.  As always,
if $r\gg1$ no spatial
structures develop; homogeneous states simply alternate in time.
On the other hand, for $r=0.94$, we again obtain oscillatory patterns,
as shown in the snapshots of Fig.~\ref{fighexagons}. It is worth
noting that this excitation density
is quite different from the one in Fig.~\ref{figsquares}. As expected,
the spots are arranged hexagonally.  Most strikingly, in
this case there is no glide oscillation but rather a true oscillation
of localized excitations whose size is again consistent with the
calculated aspect ratio. As $r$ is decreased 
the excitation field is frozen and one
obtains a stationary hexagonal pattern of spots.

We have also performed simulations of Eq.~(\ref{4}) when
the modulation $\mu(t)$ is a dichotomous noise.
In this case, the
intermediate regime of oscillatory patterns fades out and the system 
exhibits alternating homogeneous states if the correlation time of
$\mu(t)$ is large, and stable patterns (rolls or hexagons) 
if the correlation time is small~\cite{next}.

Summarizing, we have shown that the alternation of two dynamics can create
patterns even though each separate dynamics drives the system to a
homogeneous state. Moreover, we have seen that in the crossover between 
alternating homogeneous states and stationary patterns, a rich phenomenology
of oscillatory patterns may emerge.
The mechanism is very general and can easily be extended to
other situations such as reaction-diffusion systems. Thus, we 
have provided an alternative non-equilibrium
mechanism to dissipative structures or modulation of bifurcation parameters 
for the formation of stationary and oscillatory spatial patterns. 

An interesting open problem is whether it is possible to find this
type of behavior when the switching occurs between two
equilibrium dynamics. This is not the case in the SH
equation, even when the homogeneous state is stable.
For instance, in the case of convection, the homogeneous state
describes a fluid in mechanical but not in thermal
equilibrium.  If
switching between local potential is achievable in the case of
convection, our results would imply that the system will exhibit
convection patterns by alternating dynamics which separately drive the
fluid to mechanical (but not to thermal) equilibrium. 

The authors wish to thank R. Kawai for fruitful discussions. This work was
partially supported by the National Science Foundation under grant No.
PHY-9970699, by DGES-Spain Grant PB-97-0076, by the {\em New Del Amo
Program}, and by MECD-Spain Grant EX2001-02880680.

\newpage

\newpage

\newpage

\newpage

\end{document}